\documentclass[prl,superscriptaddress,showpacs,twocolumn]{revtex4}

\usepackage{graphicx}
\usepackage{amsmath}
\usepackage[nooneline,raggedright]{subfigure}

\def\vo2{VO$_2$}
\def\vcro2{V$_{1-x}$Cr$_x$O$_2$}

\def\a1g{$a_{1g}$}

\newcommand{\ttg}{$t_{2g}$}

\begin{document}

\title{Combined GW and dynamical mean field theory:\\
 Dynamical 
screening effects in transition metal oxides}

\begin{abstract}
We present the first dynamical implementation of the combined
GW and dynamical mean field scheme (``GW+DMFT'') for first 
principles calculations 
of the electronic properties of correlated materials.
The application to the ternary transition metal oxide
SrVO$_3$ demonstrates that this
schemes inherits the virtues of its two parent theories:
a good description of the local low energy correlation physics
encoded in a renormalized quasi-particle band structure, spectral
weight transfer to Hubbard bands, and the physics of screening 
driven by long-range Coulomb interactions.
Our data is in good agreement with available photoemission and inverse 
photoemission spectra; our analysis leads to a reinterpretation
of the commonly accepted ``three-peak structure'' as
originating from orbital effects rather than from the electron
addition peak within the \ttg\ manifold.
\end{abstract}

\author{Jan M. Tomczak}
\affiliation{Department of Physics and Astronomy, Rutgers University, Piscataway, New Jersey 08854, USA}
\author{Michele Casula}
\affiliation{CNRS and Institut de Min\'eralogie et 
de Physique des Milieux condens\'es,
Universit\'e Pierre et Marie Curie,
case 115, 4 place Jussieu, 75252, Paris cedex 05, France}
\author{Takashi Miyake}
\affiliation{Nanosystem Research Institute (NRI) 
RICS, AIST, Tsukuba, Ibaraki 305-8568, Japan }
\affiliation{Japan Science and Technology Agency, CREST, Kawaguchi
 332-0012, Japan}
\author{Ferdi Aryasetiawan}
\affiliation{Department of Physics, Mathematical Physics,
Lund University,
S\"olvegatan 14A,
22362 Lund, Sweden}
\author{Silke Biermann}
\affiliation{Centre de Physique Th{\'e}orique, 
Ecole Polytechnique, CNRS UMR 7644, 91128 Palaiseau
Cedex, France}
\affiliation{Japan Science and Technology Agency, CREST, Kawaguchi
 332-0012, Japan}

\maketitle


Describing electronic excitations of materials with strong electronic
Coulomb interactions remains one of the challenges of modern condensed
matter theory \cite{imada}. 
Photoemission satellite features, quasi-particle mass
renormalizations or -- in the strong coupling limit --  the localization of 
electrons are phenomena well beyond band theory. Within effective low-energy
models, such as the Hubbard or Anderson lattice model, insight has been
gained into the low-energy physics of correlated materials on a qualitative
level. However, when relating 
this knowledge to specific materials, attempting
a \textit{quantitative} description, several challenges persist:
The first one is a question of energy scales: 
in fact, much of the physics of interest in correlated
materials takes place on the scale of several eV, well beyond what is
captured within a low-energy description. 
Optical properties of transition metal oxides, to cite a specific
example, are determined by the onset of optical transitions involving ligand
states already at scales of only a few eV. 
A second issue is the determination of the model parameters, e.g., the
effective low-energy hopping parameters and Coulomb interactions, in
particular the local \textquotedblleft Hubbard $U$\textquotedblright. 
These two issues
are in fact intimately related, since high energy screening processes 
renormalize these parameters, and an appropriately downfolded
low-energy model should take this
into account
\cite{PhysRevB.70.195104,PhysRevLett.95.176405,PhysRevB.85.035115,werner_bfa,casula_effmodel,ferdi_down}.

In this Letter, we present the first \textit{dynamical}
implementation of a combined GW and dynamical mean field (``GW+DMFT'')
scheme that addresses the above issues, by treating explicitly
both, dynamical Coulomb interactions on the ``correlated'' manifold
of orbitals, and corrections to the band structure 
of degrees of freedom that live on higher energy scales.
The application to a well-studied correlated metal, SrVO$_3$, shows some expected
features: a renormalization of the quasiparticle band structure, the
formation of a lower Hubbard band and the correction of the position of the
oxygen $p$-states. It also leads to less expected features, namely the absence of a
clearly distinguishable upper Hubbard band in the total spectral function
along with a reinterpretation of the feature seen at
about 2.7 eV in inverse photoemission that we associate
with the 
e$_{g}$ states. This attribution
is in agreement with cluster model 
calculations \cite{mossanek:033104} and allows
to resolve apparent 
contradictions between cluster and dynamical mean field
methods for this compound. 
In the 
orbital-resolved spectral function we furthermore 
predict additional
satellite features induced by the frequency-dependence of the effective
Coulomb interaction. 

The combined GW+DMFT scheme was proposed a few years ago, in order to
avoid the ad hoc nature of the Hubbard parameter and the double counting
inherent to conventional combinations of dynamical mean field theory with
density functional theory in the local density approximation 
(so-called \textquotedblleft LDA+DMFT\textquotedblright\ 
schemes\cite{PhysRevB.57.6884, 0953-8984-9-35-010}).
The starting point is Hedin's GW approximation (GWA)\cite{hedin}, in which the
self-energy of a quantum many-body system is obtained 
as a product of
the Green's function G and the screened Coulomb interaction 
$W=\epsilon^{-1}V$. 
The dielectric function $\epsilon $, which screens the bare Coulomb
potential $V$, is --within a pure GW scheme-- obtained from the random
phase approximation. The GW+DMFT scheme, as proposed in \cite{PhysRevLett.90.086402}, 
combines the first principles description of screening inherent in GW
methods with the non-perturbative nature of DMFT, where local quantities
such as the local Green's function are calculated to all orders in the interaction
from an
effective reference system (\textquotedblleft impurity model\textquotedblright ).
In DMFT, one imposes a self-consistency condition for the one-particle
Green's function, namely, that its  
on--site projection equals the impurity 
Green's function. In GW+DMFT, the self-consistency requirement is
generalized to encompass two-particle quantities as well, namely, the local
projection of the screened interaction is required to equal the  
impurity screened interaction.  
This 
in principle promotes the Hubbard $U$ from an adjustable parameter in
DMFT techniques to a self-consistent auxiliary function that incorporates
long-range screening effects in an {\it ab initio} fashion. Moreover, 
the theory provides
momentum dependence
to quantities (such as the self-energy) that are local
within pure DMFT. 
Despite these conceptually attractive features, 
however, no dynamical implementation, even at the
non-selfconsistent level has been achieved so far (see however the
static implementation in \cite{PhysRevLett.90.086402},
as well as model studies of a one-band extended Hubbard model 
in \cite{PhysRevLett.92.196402,ayral_gwdmft}). 

From a formal point of view, the GW+DMFT method, as introduced in Ref.~\cite%
{PhysRevLett.90.086402}, corresponds to a specific 
approximation to the correlation 
part of the free energy of a solid, expressed as a functional of
the Green's function $G$ and the screened Coulomb interaction $W$~: the non-local part is
taken to be the first order term in $W$, while the local part 
is calculated
from
an impurity model as in (extended) dynamical mean field theory.
This leads to a set of self-consistent equations for  
$G$,  
$W$, the self-energy $\Sigma $ and the
polarization $P$ \cite{gwdmft_proc2,gwdmft_proc1}.
Specifically, the self-energy is obtained as $\Sigma
=\Sigma _{local}+\Sigma _{non-local}^{GW}$, where the local part 
$\Sigma_{local}$ is derived from the impurity model\footnote{%
The notion of locality refers to the use of a specific basis set of
atom-centered orbitals, such as muffin-tin orbitals, or atom-centered Wannier
functions.}. 
In practice, however, $s$- or $p$-orbitals are rather delocalized. 
Here, we therefore propose a practical scheme,
in which only the local part of the self-energy of the \textquotedblleft
correlated\textquotedblright\ orbitals is 
obtained from an impurity reference, 
and all other local and non-local components are
approximated by their first order expressions in $W$. 
Green's functions and self-energies are expressed as matrices 
in a basis of atom-centered Wannier functions%
\footnote{We choose a maximally localized Wannier basis\cite{PhysRevB.56.12847,miyake:085122} constructed  
separately for the correlated (\ttg) and uncorrelated subspaces, such that
the one-particle Hamiltonian is blockdiagonal. 
Then, off-blockdiagonal self-energy elements are small
and neglected in the following.
Also, the local part of the free energy functional then
separates into correlated and
uncorrelated subspace contributions. Our approach consists
in calculating only the former from the impurity model, and
the latter to first order in $W$.}.
For practical reasons, we avoid the costly GW self-consistency cycle, and in particular the
update of the dynamical interaction $U(\omega )$:
Instead, we calculate $U(\omega)$ 
from the constrained random phase approximation (cRPA)
\cite{PhysRevB.70.195104,
PhysRevB.74.125106, miyake:085122} and keep it fixed when
iterating the local one-particle quantities to self-consistency.%
\footnote{The GW calculation was performed for a 4x4x4 k-point mesh, the $t_{2g}$ part of which was Wannier interpolated
onto a 27$^3$ mesh for the GW+DMFT self-consistency.}

Our target material, SrVO$_{3}$, crystallizes in the cubic perovskite 
structure:
the V$^{4+}$ ions are surrounded by oxygen octahedra,
occupying the sites of a simple cubic
lattice. The Sr$^{2+}$ cation sits in the center of the cubes.
The electron count leaves a single $d$ electron in the V-d
states, which is largely responsible for the electronic
properties of the compound.
The octahedral crystal field splits the V-d states
into a lower-lying threefold degenerate $t_{2g}$ manifold, 
thus filled with one electron per V, and an empty $e_{g}$ doublet. 
The compound exhibits metallic resistivity with a Fermi liquid
T$^2$ behavior up to room temperature
\cite{Onoda1991281}
and temperature-independent Pauli paramagnetism without
any sign of magnetic ordering \cite{Inoue199761}. 
Hall data and NMR measurements \cite{Onoda1991281,Eisaki_phd}
as well as
angle-resolved photoemission spectra \cite{PhysRevLett.95.146404}
that find the Fermi surface of SrVO$_3$ as
cylindrical Fermi sheets, in agreement with
theory, confirm the picture of a normal Fermi liquid.

These properties make SrVO$_3$ an ideal model material
for studying the effects of electronic Coulomb interactions
\cite{PhysRevLett.69.1796,mossanek:155127, mossanek:033104, PhysRevB.78.075103, JPSJ.78.094709,
pavarini:176403, PhysRevB.72.155106, 1367-2630-7-1-188,anisimov:125119, lechermann:125120, 0953-8984-20-13-135227, 
PhysRevB.80.085101, 0953-8984-23-8-085601,amadon:205112}. 
Recent works include, e.g., studies of
surface effects \cite{PhysRevLett.90.096401}, 
kink structures in the momentum-resolved
spectral function \cite{nekrasov:155112,byczuk-2007-3,PhysRevLett.109.056401},
or non-local correlation effects \cite{PhysRevB.76.045108,PhysRevB.85.165103}
in relation to laser ARPES \cite{PhysRevLett.96.076402} 
which finds a ``dip'' at the Fermi 
level or a maximum of the quasi-particle peak slightly
below (at around -0.2 eV). 
The experimentally observed quasiparticle weight is
about 0.5-0.6 \cite{PhysRevLett.69.1796,
Inoue_phd, PhysRevB.58.4372, 0295-5075-55-2-246,
PhysRevB.73.052508,
PhysRevLett.93.156402, PhysRevB.80.235104,PhysRevB.82.085119}, 
with the remaining spectral weight being
carried by a photoemission (PES) (Hubbard-)satellite 
at around -1.6 eV 
\cite{PhysRevB.52.13711,PhysRevLett.93.156402}. 
Inverse photoemission (BIS) has located the
electron addition $d^1 \rightarrow d^2$ peak at an energy of about
2.7 eV \cite{PhysRevB.52.13711}.

\begin{figure}[!hb]   
\subfigure[\, ]{\scalebox{1.}{\includegraphics[clip=true,trim=15 10 500 5,angle=-90,width=.44\textwidth]{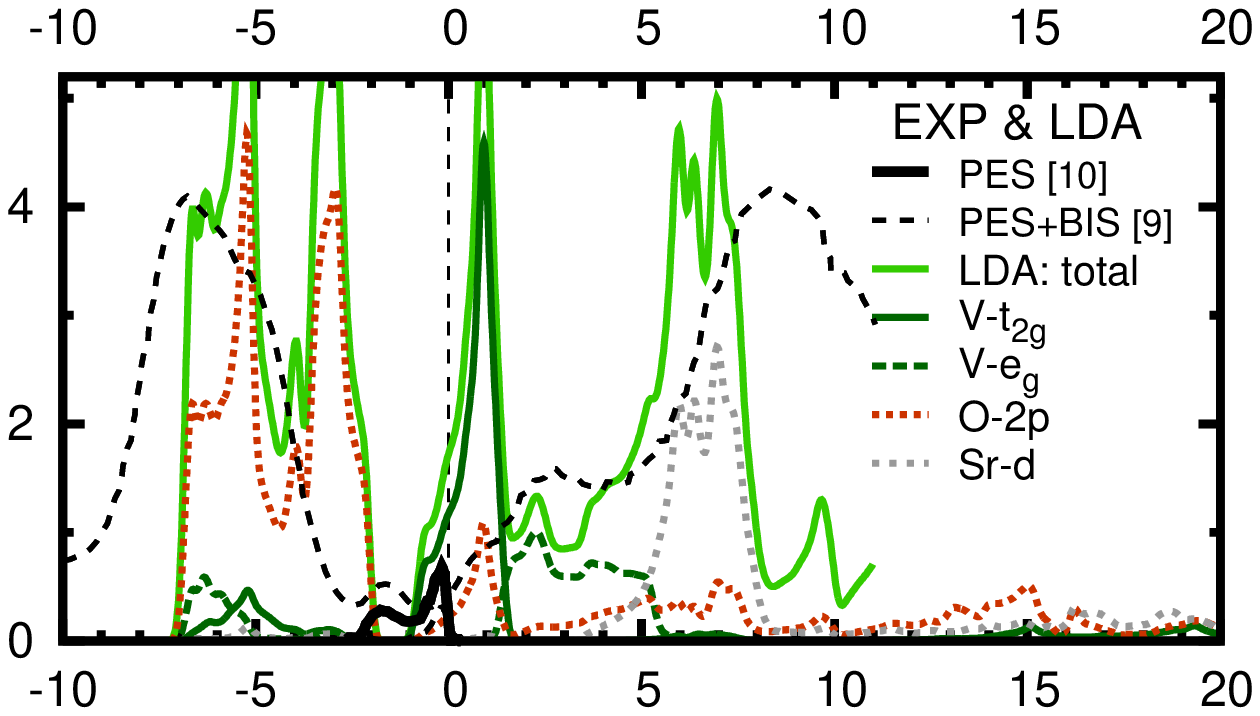}}}
\subfigure[\, ]{\scalebox{1.}{\includegraphics[clip=true,trim=175 10 60 5,angle=-90,width=.44\textwidth]{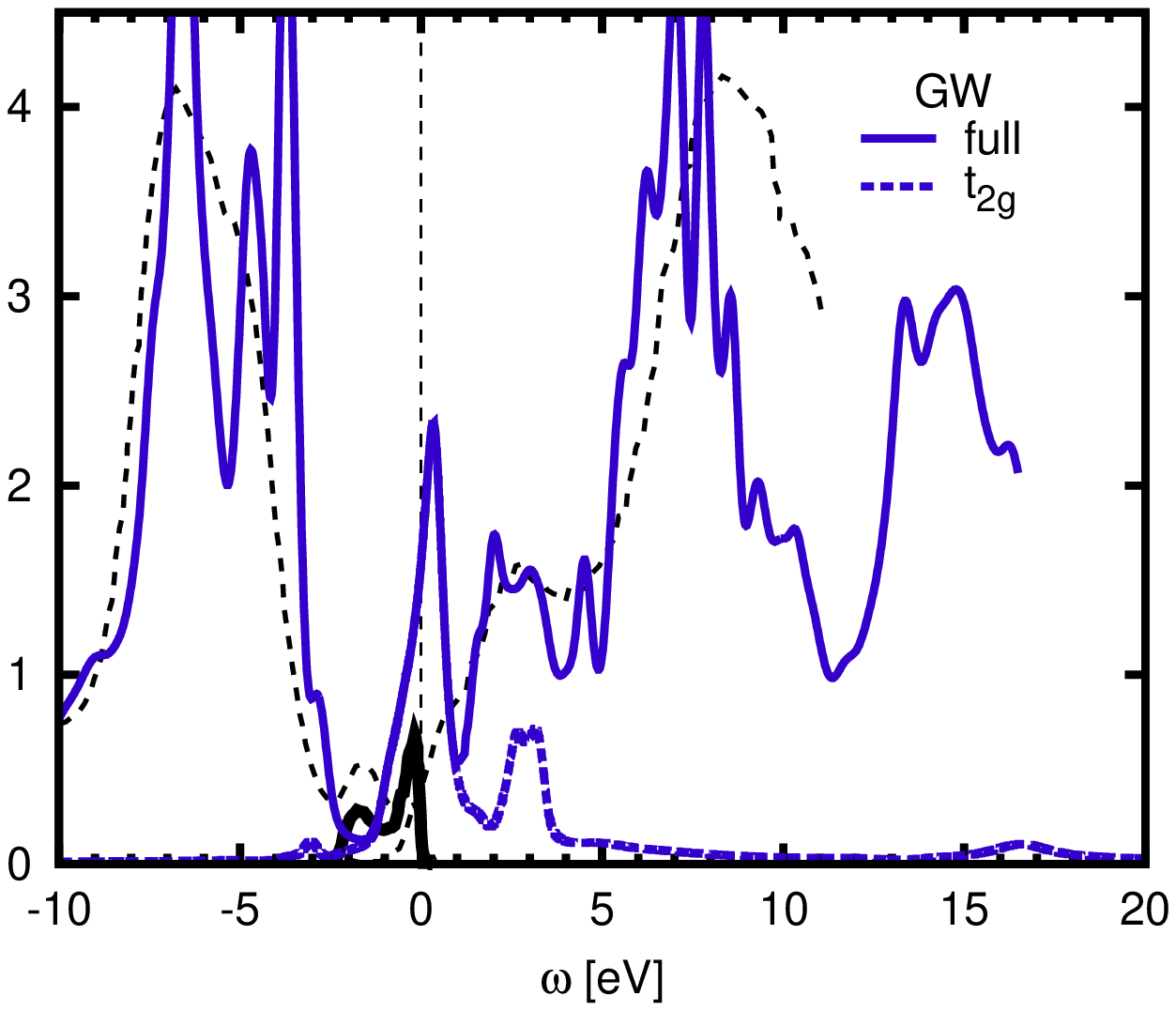}}}
\subfigure[\, ]{\scalebox{1.}{\includegraphics[clip=true,trim=175 10 60 5,angle=-90,width=.44\textwidth]{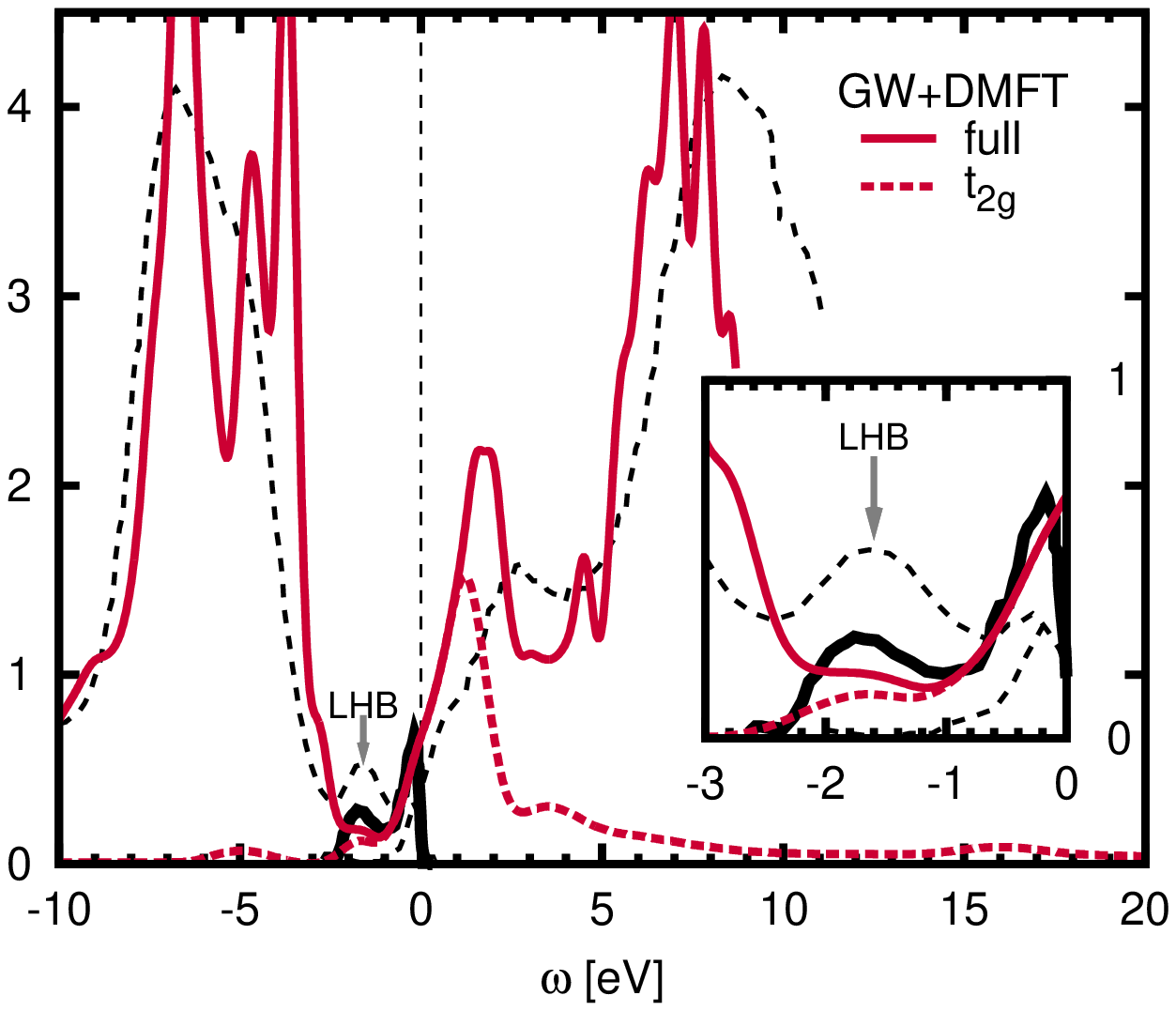}}}
\caption[bla]{
(a) Orbital-resolved LDA density of states, in comparison to experimental
spectra from photoemission (PES)\cite{PhysRevB.52.13711,PhysRevLett.93.156402} and inverse photoemission (BIS)\cite{PhysRevLett.93.156402}.
The latter are reproduced also in panels (b-c).
(b) Spectral function from GW (full orbital and $t_{2g}$ only) in comparison to experiments.
(c) Spectral function from GW+DMFT (full orbital and $t_{2g}$ only) in comparison to experiments.
The inset shows a low energy zoom. 
In the GW+DMFT spectrum, LHB denotes the lower Hubbard band. 
See text for details.
}
\label{fig:rho}
\end{figure}

Figure~\ref{fig:rho}a summarizes the LDA electronic structure:
the O-p states disperse between -2 and -7 eV, separated from the
\ttg\ states whose bandwidth extends from -1 eV to 1.5 eV.
While the \ttg\ and e$_{g}$ bands are well separated at every given k-point,
the partial density of states (DOS) slightly overlap, and the e$_{g}$ states display a pronounced
peak at 2.3 eV. Finally, peaks stemming from the Sr-d states are
located at  6.1 eV and 7.1 eV.
We have superimposed to the LDA DOS the experimental PES and BIS 
curves taken from \cite{PhysRevB.52.13711,PhysRevLett.93.156402}. The comparison reveals the main
effects of electronic correlations in this material: as expected on 
quite general grounds, LDA locates the filled O-p states at too high 
and the empty Sr-d manifold at too low energies.
The \ttg\ manifold undergoes a strong quasi-particle
renormalization with a concomitant shift of spectral weight to the lower
Hubbard band, both of which are effects beyond the one-particle
picture.

Applying the GW approximation (see the spectral function in Fig.~\ref{fig:rho}b)
increases the O-p to Sr-d distance, placing both manifolds at
energies nearly in agreement with experiment\footnote{%
The persisting slight underestimation is expected~: (a)
the polarization (determining $W$) is derived from
the bare t$_{2g}$ bands, whose character
is largely too itinerant in LDA, and (b) vertex corrections are neglected.   
}.
Most interestingly, however, a peak at 2.6 eV emerges from the
$d$-manifold, which we find to be of e$_{g}$ character. Indeed, the
GW approximation enhances the crystal field splitting and
places the maximum of the e$_{g}$ spectral weight at the location of
the experimentally observed $d^1 \rightarrow d^2$ addition peak.

\begin{figure}[b!]
\includegraphics[clip=true,trim=35 15 50 15,angle=-90,width=0.49\textwidth]{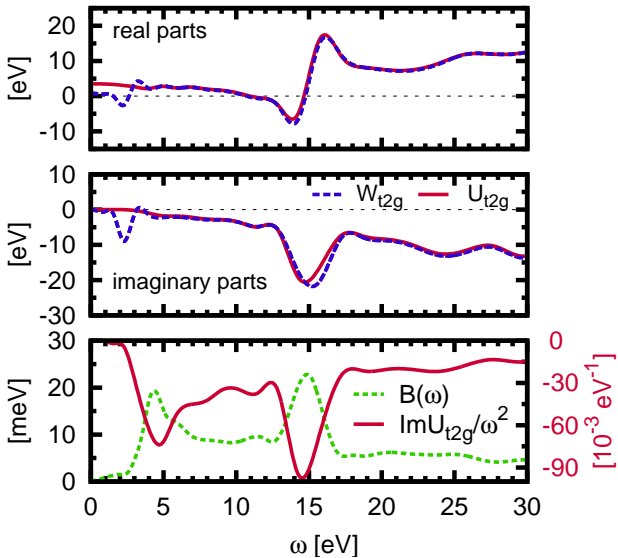}
\caption[bla]{(Partially) screened Coulomb interactions for SrVO$_3$:
$\mathrm{Re} U$, $\mathrm{Re} W$ (top panel),   $\mathrm{Im} U$, $\mathrm{Im} W$ (middle).
$\mathrm{Im} U/\omega^2$ and plasmon spectral function $B(\omega)$ (bottom).
}
\label{fig:W}
\end{figure}

Also displayed is the partial \ttg\ contribution. 
Two interesting points are seen~: the quasi-particle width is
narrowed from the LDA value by about 0.5 eV and satellite 
structures appear below and above the Fermi level
at $\pm$3~eV, $+$5eV as well as at around 15eV. 
Within the (non-self-consistent) GW approximation, the self-energy 
is related to $\mathrm{Im} W$ as follows~:
\begin{eqnarray*}
\mathrm{Im} \Sigma _{GW}(r,r^{\prime };\omega ) 
&=& \sum_{kn} \psi _{kn}(r)\mathrm{Im} W(r^{\prime },r;
\varepsilon_{kn}-\omega )\psi _{kn}^{\ast }(r^{\prime })
\end{eqnarray*}%
Thus, a feature at $\omega_0$ in $\mathrm{Im}W$ causes a satellite feature at an
energy $\omega_{0}$ below or above the quasiparticle peak at $\varepsilon_{kn}$.
In Fig.~\ref{fig:W} we display the
on-site \ttg\ intra--orbital element $W_{t2g}$ of the fully screened interaction in the Wannier basis.
Three notable features are clearly discernible: There is the usual high-energy
electron-gas-like plasmon excitation at around 15 eV 
and, more remarkably,
there is a strong excitation between 2--3
eV, as well as a shoulder at $\sim$4.5eV.
Evidently, these energy positions directly correlate with the spectral features discussed above.
It is important to note that the lowest feature in $W_{t2g}$
mainly arises from a collective excitation within the 
$t_{2g}$ bands.
This  can be understood by comparison with the 
Hubbard interaction $U(\omega)$ of the low energy 
\ttg\ system, as calculated within the constrained RPA.
In $U(\omega)$ the structure at 2$\sim$3
eV is absent, due to the elimination of the polarization within the $t_{2g}$ bands.  
The shoulder at around 4.5eV, however, partly persists, as its excitations also involve O-2p states;
we will come back to this point below.

Solving the GW+DMFT equations, even without the self-consistency
cycle at the GW level, that is, for fixed $\Sigma _{GW}^{non-local}$ and $U(\omega)$, is a hard task. We
use the scheme recently introduced in Ref.~\cite{PhysRevB.85.035115}:
the Green's function of the dynamical impurity model is
obtained from a factorization ansatz
$G(\tau )=G_{stat}(\tau)B(\tau)$ where $G_{stat}$ is the Green's 
function of a static impurity model with constant $U$$=$$U(\omega$$=$$0)$%
\footnote{For the Hund's rule coupling the static cRPA value $J=0.5$eV is used.},
and the second factor $B(\tau)=G(\tau )/G_{stat}(\tau )$
is approximated by its
value for vanishing bath hybridization $\Delta$\cite{PhysRevB.85.035115} :
\begin{equation}
B(\tau) = \exp\left({- \int_0^{\infty} \frac{d\omega}{\pi}
\frac{\mathrm{Im} U(\omega)}{\omega^2}\left[ K_{\tau}(\omega) - K_{0}(\omega)\right]}\right)
\label{factorisation-approximation} 
\end{equation}%
with $K_{\tau}(\omega)=\frac{\exp(-\tau \omega) + \exp(-(\beta- \tau) \omega)}
{1 - \exp(-\beta \omega)}$.
In the regime that we are interested in, namely 
when the plasma frequency is
several times the bandwidth, this is an excellent approximation\cite{PhysRevB.85.035115}. 

\begin{figure}[t!h]
\includegraphics[clip=true,trim=10 15 45 10, angle=-90,width=0.49\textwidth]{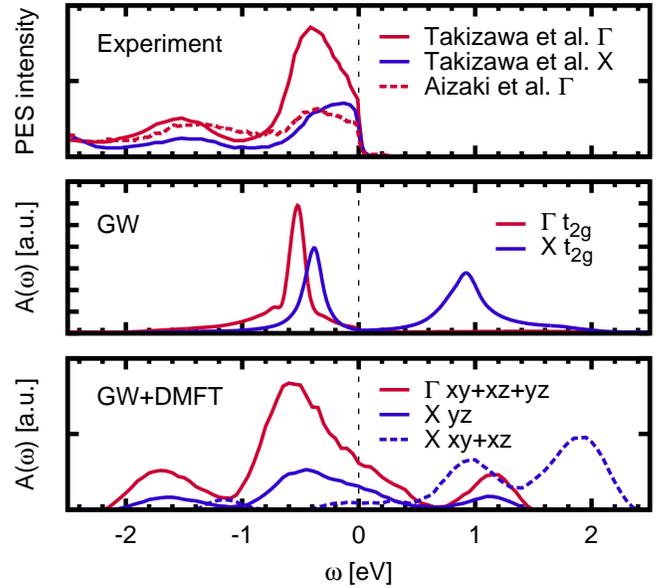}
\caption[bla]{Momentum-resolved \ttg\ spectral functions  
from GW and GW+DMFT, compared to experimental angle resolved photoemission
spectra of Refs.~\cite{PhysRevLett.109.056401} 
and \cite{PhysRevB.80.235104}.
The GW spectra are calculated at zero temperature.
}
\label{fig:arpes} 
\end{figure}

Finally, we present our GW+DMFT results for the spectrum of SrVO$_{3}$ at inverse temperature $\beta$$=$$10$eV$^{-1}$~:
Fig.~\ref{fig:rho}(c) displays the local projection, while Fig.~\ref{fig:arpes} shows 
momentum dependent \ttg\ spectra in comparison with ARPES measurements\cite{PhysRevB.80.235104,PhysRevLett.109.056401}. 
The low-energy part of the spectrum is dominated by
the \ttg\ contribution, which, here, is profoundly modified with respect to pure GW results.
A renormalized
quasi-particle band disperses around the Fermi level~: At the $\Gamma$ point the peak is located at about -0.5 eV --
a strong renormalization of the corresponding Kohn-Sham state which, at this momentum, has an energy of -1 eV.
At the X-point, the \ttg\ bands are no longer degenerate,
and surprisingly weakly renormalized xy/xz states are observed
at 0.9 eV, while the yz band is located at nearly the
same energy as at the $\Gamma$ point, in agreement
with ARPES.  
At binding energies of 1.6 eV, ARPES witnesses a weakly dispersive
Hubbard band, whose intensity varies significantly as a function 
of momentum \cite{PhysRevB.80.235104}. 
In the GW+DMFT spectral function the Hubbard band -- absent in GW -- is correctly observed
at about -1.6 eV,  
and its k-dependent intensity
variation (see Fig.~\ref{fig:arpes}) is indeed quite strong.
Previous LDA+DMFT calculations placed 
the lower Hubbard band at larger negative energies (see e.g.\ \cite{pavarini:176403}).
This is owing to the fact that when using a static Hubbard interaction, a value of 4--6 eV,
that is 
larger than the zero frequency limit of the {\it ab intio} $U(\omega$$=$$0)$$=$$3.6$eV\cite{miyake:085122}, is needed to account
for the observed transfers of spectral weight.
LDA+$U(\omega)$+DMFT
calculations taking not only the {\it ab initio}
value of the static component of $U$ but 
also its full frequency dependence into account
indeed reproduce the position of the
lower Hubbard band quite well \cite{PhysRevB.85.035115,2012arXiv1205.6965H}.
Analogously, in GW+DMFT, the additional transfers from the dynamical screening 
\cite{PhysRevB.85.035115,casula_effmodel,PhysRevLett.104.146401},
yield a good description of the Hubbard band and the spectral weight reduction at the same time.

At positive energies
non-local
self-energy effects are larger.
Interestingly, our total spectral function,
Fig.\ref{fig:rho}(c), does not display
a clearly separated Hubbard band. The reason is visible from the 
k-resolved spectra: the upper Hubbard band is located at around
2 eV, as expected from the location of the lower Hubbard band and
the fact that their separation is roughly given by the 
zero-frequency value of $U$. 
The peak around 2.7 eV that appears in the inverse photoemission
spectrum \cite{PhysRevB.52.13711} -- commonly interpreted as the
upper Hubbard band of \ttg\ character in the DMFT literature -- 
arises in fact from e$_{g}$ states located in this energy range.
The non-local
self-energy effects lead, in the unoccupied part of the
spectrum, to overlapping features from different k-points and
an overall smearing of the
total spectral function.

Akin to the correspondence between $\mathrm{Im}W$ and the GW spectrum, our approach to solve the GW+DMFT equations
allows for a transparent physical interpretation of the arising spectral properties.
Indeed the spectral representation of the bosonic renormalization factor $B(\tau)$ of Eq.~\ref{factorisation-approximation}
(displayed in the lower panel of Fig.~\ref{fig:W}) 
is directly related
to the density of screening modes
$\mathrm{Im} U(\omega)/\omega^2$ \cite{PhysRevB.85.035115}. 
In this way, we can trace back the GW+DMFT satellite 
at -4.5 eV 
to the onset of $p$-\ttg\ excitations,
discussed above for $W$ and $U$. 
On the other hand, since the feature below 3~eV in $W$ is
absent in $U$ and $B$, 
at least the local self-energy contribution to the corresponding peak within GW
does not survive in GW+DMFT.
The strong peak at 15 eV is the 
well-known plasma excitation, seen
e.g.\ in electron energy loss spectra of SrTiO$_3$\cite{PhysRevB.62.7964}.

In conclusion, we have presented the first dynamical implementation
of a combined GW and dynamical mean field scheme. 
Quite generally, materials with a
\textquotedblleft double LDA failure\textquotedblright --
an inappropriate description of 
correlated states, and deficiencies of LDA for the
more itinerant states, such as an underestimated
``$pd$-gap'' -- can be treated within our scheme. 
Application to
a well-studied benchmark material, SrVO$_3$, confirms the ability
of the approach to describe simultaneously Hubbard bands, higher energy
satellite structures and corrected energy gaps, in an {\it ab initio} 
fashion.
In the unoccupied part of the one-particle
spectral function, we attribute a strong peak evidenced in BIS
spectra to the electron addition process into the e$_{g}$ states,
in agreement with cluster studies \cite{mossanek:033104}. 
We further predict the existence of a satellite feature at -4.5 eV, that
may seem difficult to resolve experimentally  because of its
 overlap with oxygen $p$-states. 
Nevertheless we would expect
that a systematic PES study as a function of photon energy, due to the
dependence of the cross section on orbital characters, could possibly
resolve this feature. In fact, structures of similar origin may
already have been observed in other materials,
such as e.g.\ in  VO$_2$ \cite{eguchi_vo2} and
iron pnictides \cite{werner_bfa}.

\acknowledgments
We thank A. Fujimori for useful discussions and for
making the data of Ref.~\cite{PhysRevLett.109.056401} available to us prior to publication.
This work was supported by the French ANR under project
SURMOTT, and a grant of supercomputing time at IDRIS/GENCI
under project number 1393.


\end{document}